# SECONDARY ION MASS SPECTROMETRY FOR SRF CAVITY MATERIALS

J. Tuggle[1*], U. Pudasaini[2], A.D. Palczewski[3], C.E. Reece[3], F.A. Stevie[4], M.J. Kelley[1,2,3]
[1]Virginia Polytechnic Institute and State University, Blacksburg, VA, USA
[2]The College of William and Mary, Williamsburg, VA, USA
[3]Thomas Jefferson National Accelerator Facility, Newport News, VA, USA
[4]Analytical Instrumentation Facility, North Carolina State University, Raleigh, NC, USA
Corresponding Author: JaysMail@vt.edu

*Abstract*

Historically, many advances in superconducting radio frequency (SRF) cavities destined for use in advanced particle accelerators have come empirically, through the iterative procedure of modifying processing and then performance testing. However, material structure is directly responsible for performance. Understanding, the link between processing, structure, and performance will streamline and accelerate the research process. In order to connect processing, structure, and performance, accurate and robust materials characterization methods are needed. Here one such method, SIMS, is discussed with focus on analysis of SRF materials. In addition, several examples are presented, showing how SIMS is being used to further understanding of materials based SRF technologies.

## INTRODUCTION

### *Motivation for Work with SRF Materials*

Super conducting radio frequency particle accelerators are invaluable tools in the push to expand the frontier of scientific understanding. In order to push further, accelerators with higher particle energies, increased beam current, and reduced cost per unit performance must be constructed. Accelerator performance is chiefly controlled by the superconducting niobium cavities at their heart, most especially by the ~40 nm rf-active surface layer at the cavity interior. Present technology pays much attention to niobium purity, measured by residual resistance ratio (RRR), and governed by bulk composition specification in the ppm range for interstitials elements (C, N, O). How this translates to composition of the active layer and in turn into performance is not known, as the composition of the active layer is not measured.

Until recently (~2013) it was believed elements such as H, C, O and N were contaminants in SRF Nb to be avoided at high cost. [1, 2, 3, 4, 5] However, extensive evidence from cavity processing and testing indicates that introduction of small amounts of nitrogen and perhaps other interstitials can markedly improve energy efficiency, as characterized by quality factor $Q_0$. [6, 7, 8, 9, 10] Difficulties encountered in current efforts to implement doping technology for the Linac Coherent Light Source II (LCLS II) suggest that deeper understanding is needed. [11, 12]

Another potential path to improvement is the creation of a micron-thick layer of $Nb_3Sn$ on the cavity interior surface. [13] The viable operating temperature for many applications moves from 2 K (superfluid helium) to 4.2 K (liquid helium), with great savings in cryogenics expense and complication. Further, the maximum theoretical surface magnetic field (which scales with accelerating gradient) is approximately doubled. [14, 13] However, the quality factor for current coated cavities typically falls with increasing gradient so significantly that the technology has not been usefully implemented in accelerators. It is not clear why this occurs nor why, in a few instances, it has not occurred. [15]

A common thread for these and other issues is the need for better understanding of how composition and processing affect the active layer and how that in turn affects performance. A major aspect is composition measurements down to a few tens of ppm at a dimensional scale down to a few nm. Only Secondary Ion Mass Spectrometry (SIMS) has this capability.

### *Secondary Ion Mass Spectrometry*

SIMS originated in the 1970's and has since found wide use in both industrial and research institutions. This is in large part because, of all analytical techniques, SIMS has the lowest detection limits, down to 1 ppb in ideal conditions, and the ability to detect all elements.

Since the 70's the general principles of operation have remained the same. A primary ion beam is focused by a series of electrostatic lenses and used to bombard the sample surface of interest. This primary ion beam may consist of many types of ions depending on the goals of analysis, in general the most common are oxygen ions ($^{16}O_2^+$, $^{16}O^-$) or cesium ($^{133}Cs^+$). The primary beam bombards the sample at a high enough energy to cause the ejection of sample material. A small amount of this material is ejected as ions, and extracted as a secondary ion beam. The mass distribution and intensity of the ejected (secondary) ions can then be measured, from which is inferred the composition of the source. The relationship may be notionally stated as:

$$I_{sec} = I_{prim} Y_{sec} S_{sec} D_{sec} C_x$$

Where $I_{sec}$ is the measured ion intensity (counts/sec) of ions for the particular secondary ion, $I_{prim}$ represents primary ion beam condition (energy and current); $Y_{sec}$, $S_{sec}$ and $D_{sec}$ are respectively the ion yield (ions per sputtered atom), the sputter yield (sputter atoms per bombarding ion) and instrument sensitivity for the particular species under the chosen conditions. $C_x$ represents the concentration of the species of interest in the sample.

When considering $I_{prim}$, conditions are chosen in order to maximize $I_{sec}$ while still meeting target spatial resolution and detection limits. This harmonizing of $I_{sec}$ and $I_{prim}$ must consider various trade-offs. For example, SIMS operates by sputtering material from the sample surface and enough must be sputtered that adequate counts are accumulated in the particular species' peak to provide the measurement. Measuring a specific dopant level requires accumulating a certain number of counts over background at the peak position of the particular ion by sputtering a corresponding amount of material. So, sputtering a larger area permits sputtering to a shallower depth: lateral resolution trades off against depth resolution. Hence, SIMS instruments cannot be optimized simultaneously for spatial resolution and sensitivity. [16] Lateral resolutions can vary from ~50 nm, making it possible to image grain boundary segregation, to hundreds of microns when trace quantification is needed. [17, 18]

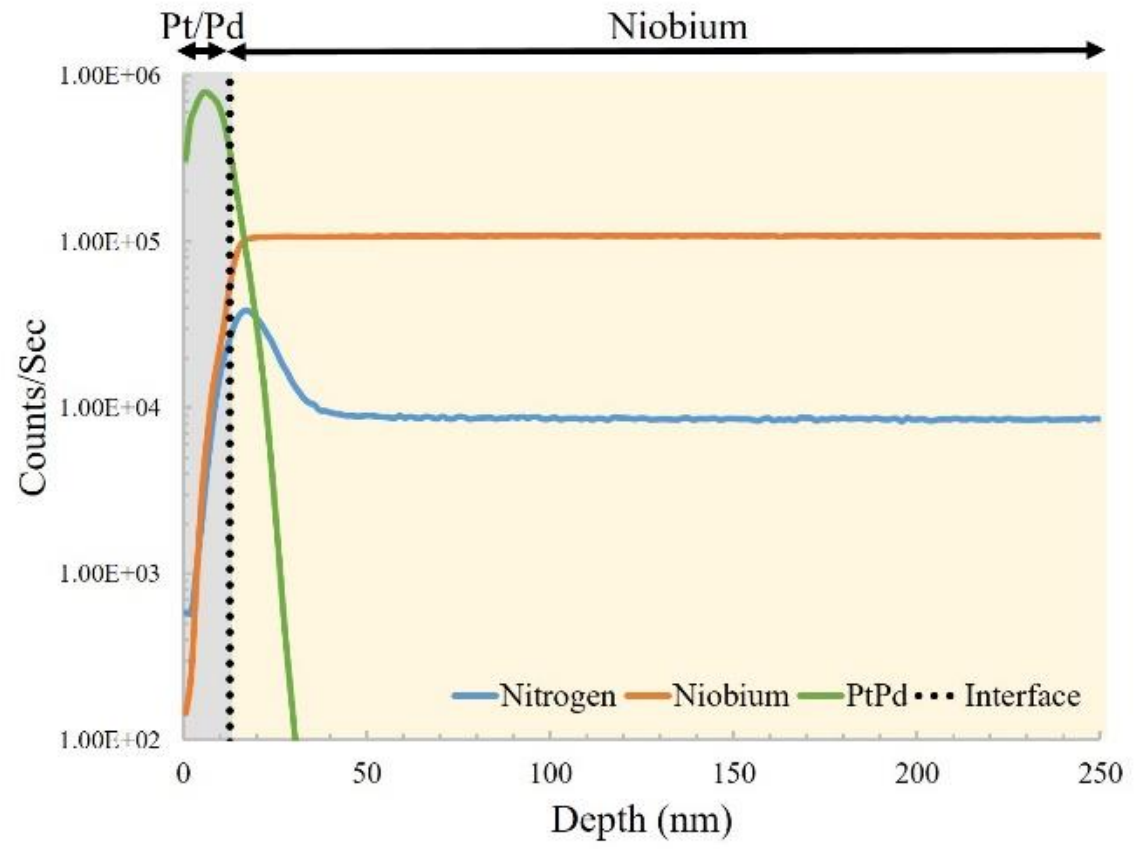

Figure 1: Depth profile of N-doped niobium with EP surface. Specimen was sputter coated with Pt/Pd prior to depth profiling.

For bulk nitrogen measurements in raw materials or N-doped witness coupons, the highest priority for a SIMS analysis method is detection limit and proper quantification. However, with the RF penetration depth controlling cavity performance, measuring nitrogen concentration near the surface (≤40 nm) with relatively high depth resolution is of concern. TOF-SIMS instruments are particularly good at high depth resolution and shallow analysis, with depth resolution less than 1 nm in ideal cases. [19] However, dynamic SIMS is capable of better detection limits and sufficient depth resolution. With the need to analyze near surface N, depth resolution must be determined under conditions that provide a sufficient sensitivity.

For an interface, depth resolution can be given as the change in depth between 84% and 16% of the maximum signal. [20, 21] In order to measure depth resolution, an interface was created by sputter coating a N-doped, electropolished (EP) witness sample with ~15 nm of platinum and palladium (80/20 by weight). This created an interface in the resulting depth profile (Fig. 3) which could then be used to calculate the depth resolution. The depth resolution was calculated using the leading and trailing edge of the interface and found to be 6.0 and 6.4 nm respectively. Figure 1 shows a depth profile of the Pt/Pd coated sample with the physical location of the interface marked by the dashed line.

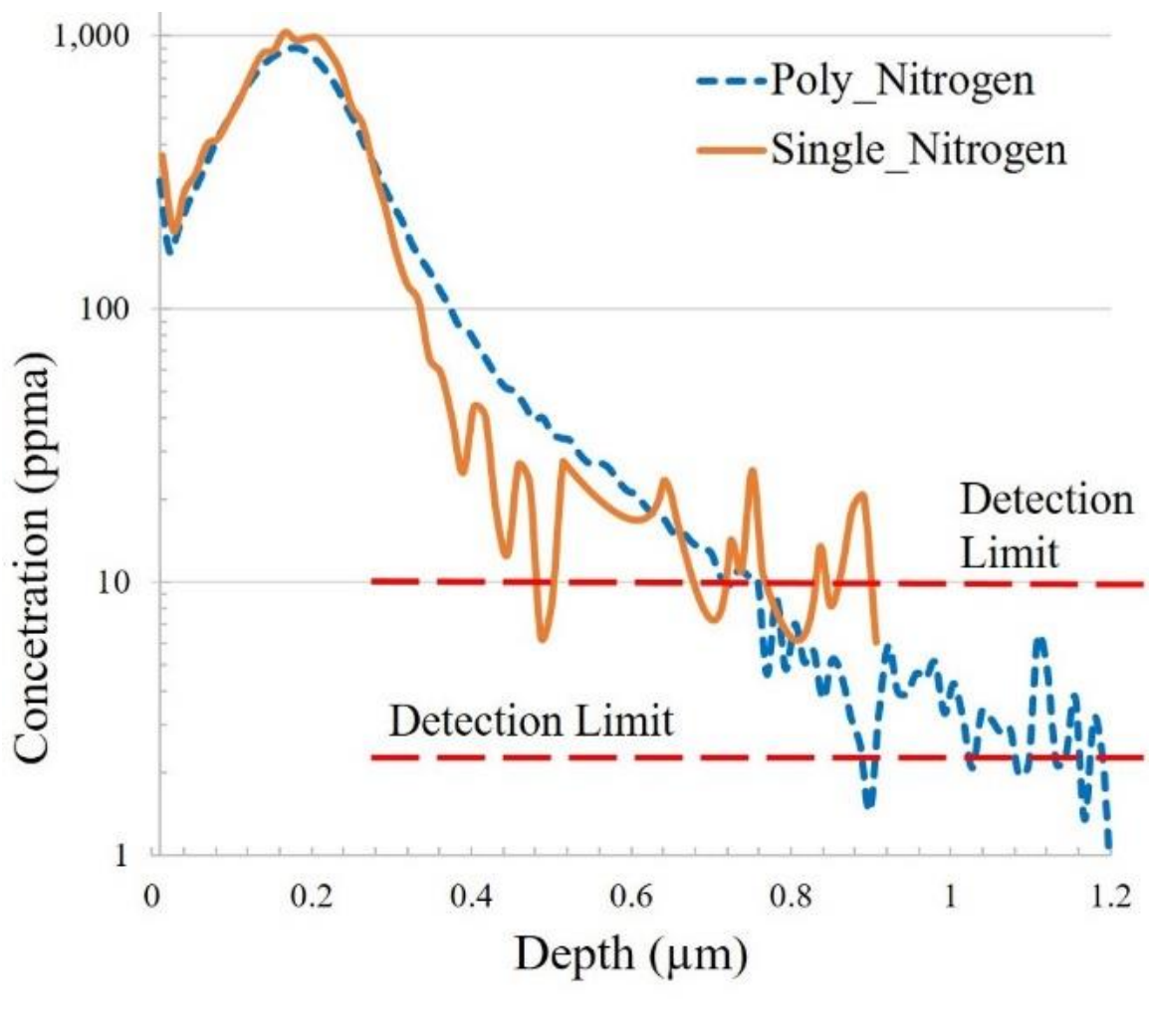

Figure 2: Depth profile of nitrogen implants used to calibrate nitrogen for both polycrystalline and single crystal material.

There are multiple sample-dependent factors which affect the secondary sputter yield ($S_{sec}$). Differing grain orientation is one which can cause sputter rates to vary greatly and negatively affect reproducibility and depth resolution. For example, while both implants in Figure 2 were created and analyzed under the same conditions, the single crystal depth profile shows a sharper implant peak and quicker drop to detection limit, indicative of better depth resolution. Depth resolution was estimated using roughness measurements and TRIM calculations and found to improve from ~109 nm for the polycrystalline implant to ~12 nm for the single crystal implant.

Surface topography will also negatively affect depth resolution and repeatability. [20] As a crater is sputtered, the original surface topography carries down and may increase with time. Normal niobium BCP surface finishes are insufficiently smooth and exhibit poor depth resolution and repeatability. Figure 3A shows multiple analyses from a single N-doped sample with BCP surface finish. In contrast, nanopolished (NP) samples have been found to reduce surface roughness to only a few nanometers and exhibit excellent repeatability in nitrogen measurements. [22] Figure 3B shows analyses from two NP samples, which were doped under different conditions. The difference in doping result can be clearly seen using NP samples. However, the error in the single BCP sample was larger than the difference seen, meaning the resulting differences in N concentration could not have been distinguished using normal BCP witness samples.

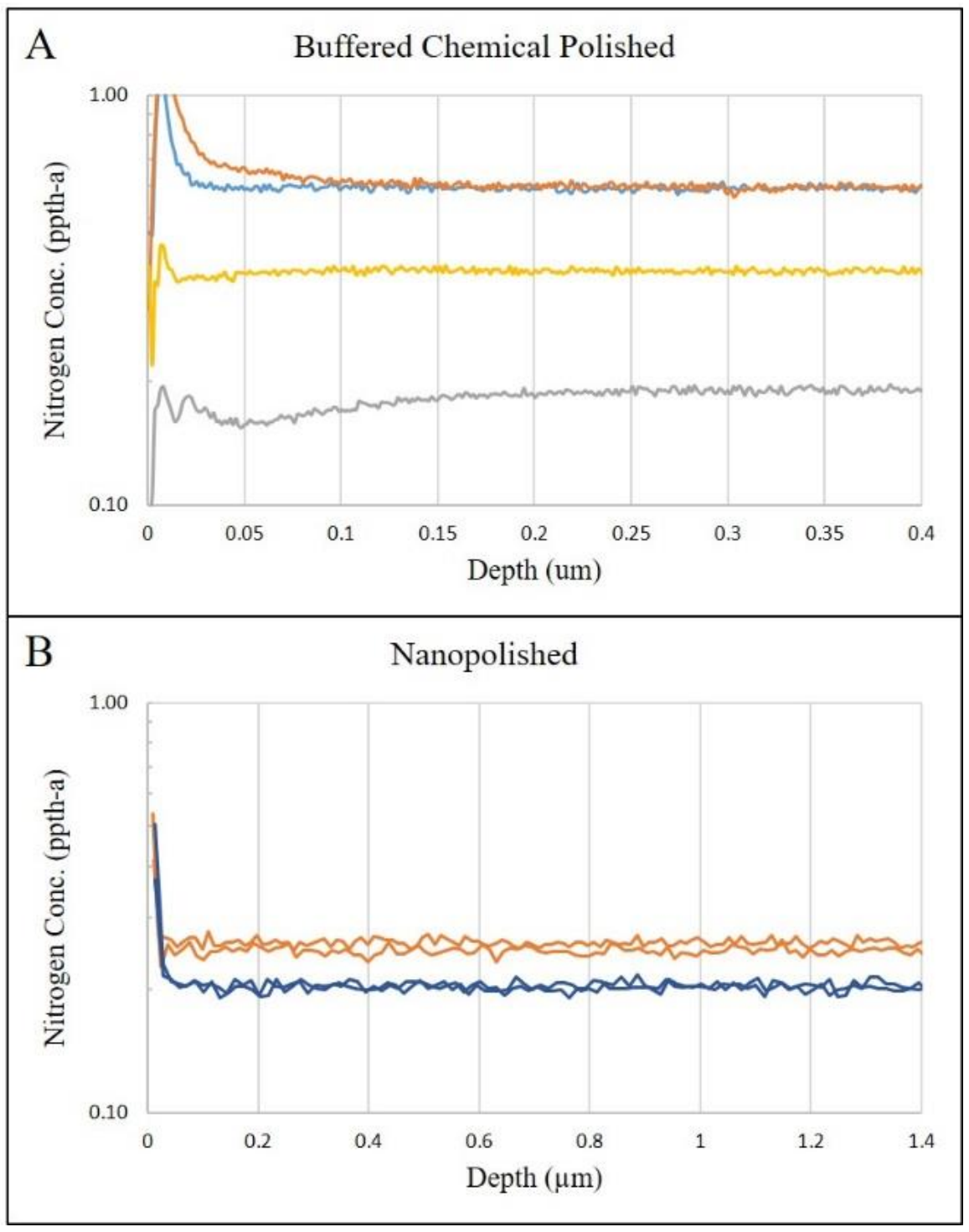

Figure 3: A and B show depth profiles with measured nitrogen concentration (parts per thousand atomic) from N-doped BCP and NP samples.

SIMS can be used to analyze species over many orders of magnitude from 100% to a lower detection limit on the order of 1 ppb. For low concentration analysis it is imperative to have low background signal from the species of interest in the instrument. This requires paying special attention when the species of interest are atmospherics such as oxygen and nitrogen. Oxygen and nitrogen are a large component of the surrounding environment and can be difficult to eliminate from a high vacuum environment. For experiments in which nitrogen and oxygen were of interest, analyses of samples were not started until an instrument vacuum of ~$3\times10^{-10}$ Torr was achieved. In addition, a cold probe surrounding the sample and cooled via liquid nitrogen was used to condense any gas directly surrounding the sample in vacuum.

Analysis of elements at low concentration, requires insuring the detection limit of the method and instrumentation is acceptable; i.e., lower than the subject species concentration. Implant standards, can be used to determine detection limits for a given analysis at the existing vacuum and instrument condition. The detection limit at time of analysis is marked on Figure 2. The vacuum condition of the instrument was slightly worse during the single crystal analysis shown here and a significant loss in detection limit is observed.

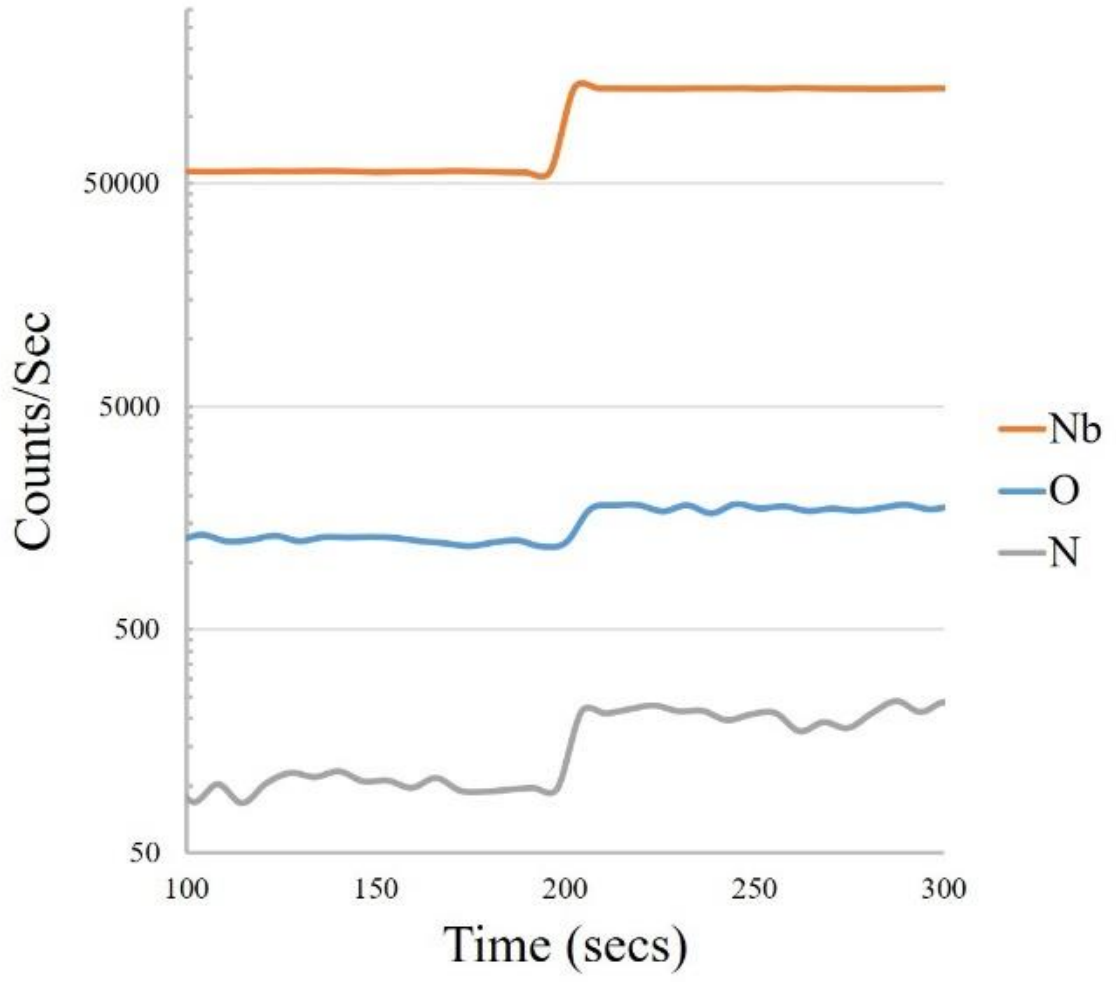

Figure 4: Raster reduction to check source of the ions detected for nitrogen and oxygen.

Because atmospherics, such as nitrogen and oxygen, are ever present in some amount, it can be difficult to know whether a baseline value for a sample is due to the species of interest within the sample or simply background instrument contamination. One way to empirically determine the nature of a detected signal is by conducting a raster reduction test. First a matrix level signal and impurity species of interest are collected at a larger raster. The raster is then collapsed while keeping the beam current and analysis area constant; 100 nA and 63 µm were used for the example here. If the raster is collapsed from 250 × 250 µm to 150 × 150 µm, as was the case in Fig. 4. This will cause the matrix signal to increase by a factor relative to the change in raster size. In the case of our example, it will increase by a factor of approximately three. If the species of interest increases this same amount, then the detected signal originates from the sample, with little background input. If the species of interest increases, but less than the matrix level signal, then there is some impurity detected from the sample, but it is near the detection limit. If the signal does not change, then no impurity species was detected and the detection limit is at least a factor of three better than before raster collapse.

The "as received" materials, discussed later, have relatively low concentrations of both N and O, making it necessary to check the origin of the species. A raster reduction was done each day that "as-received" samples were analyzed. In Fig. 4 we can see the raster reduction from an analysis day, the nitrogen detected is still well above the baseline for the method/instrument, proving the nitrogen signal measured is from the sample. The oxygen signal however appears to be approaching the detection limit and has some contribution from outside the sample.

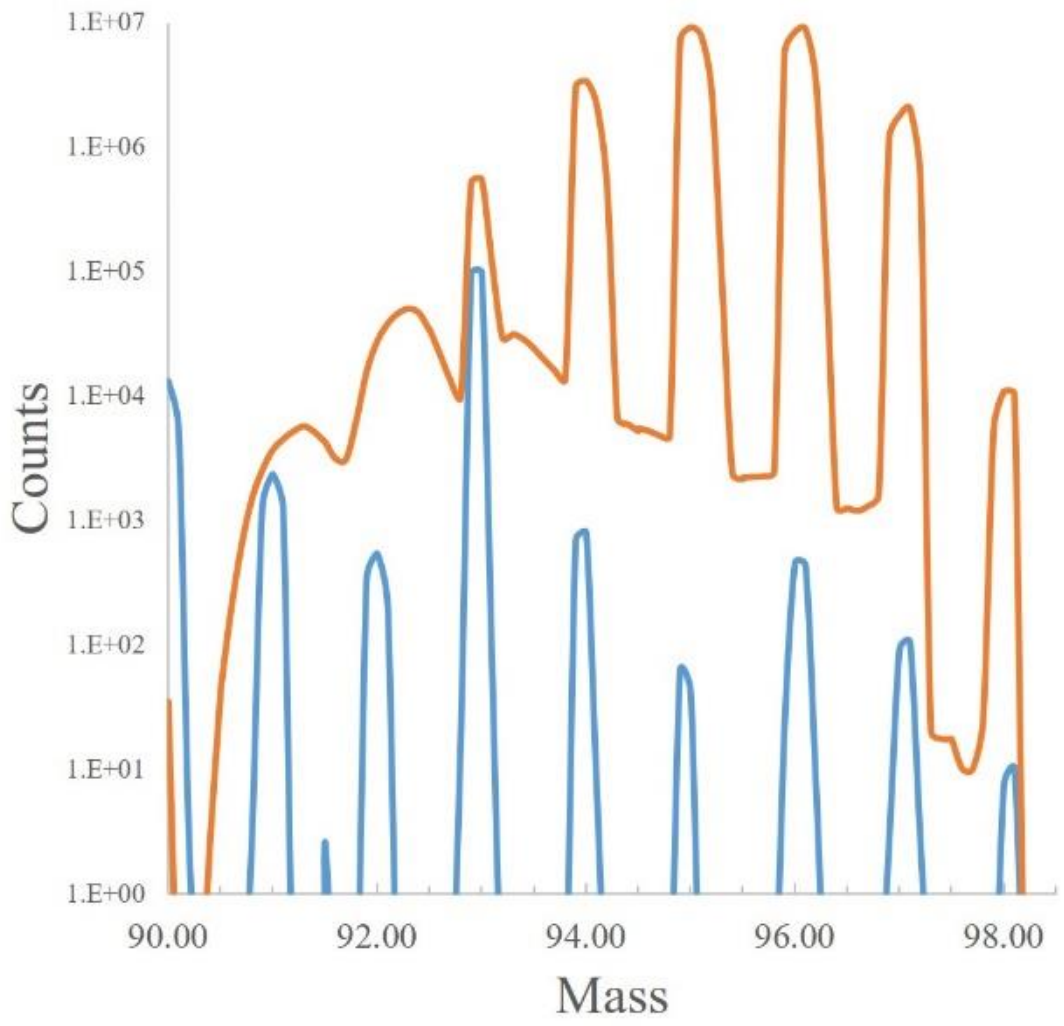

Figure 5: Mass spectra, one showing a sample with clearly resolved peaks (blue) and one suffering from metastable hydride interference (orange).

While it is useful to discuss concentration measurements in terms of the factors above to gain insight into SIMS, accurate quantification is non-trivial, requiring standards and SIMS experiments as opposed to straight forward analysis common to other analytical techniques. This is due in large part to "matrix effects". Secondary ion yields can vary five or six orders of magnitude across the periodic table and also several orders of magnitude depending on the bulk material (matrix). For an element of interest, a species is chosen which maximizes $Y_{sec}$. Nitrogen's, negative ion yield is virtually zero under many analysis conditions, requiring detection of molecular species ($^{93}Nb^{14}N^{-}$ used here) to quantify nitrogen concentration. [21] Due to variation in $Y_{sec}$, quantification cannot be based on relative signal intensities only.

The most common method for quantifying SIMS depth profiles is by utilizing ion implant standards. An implant standard must be created for each species of interest by implanting it into the matrix of interest, preferably near the same concentration range. Its depth profile can then be acquired and dose information can be used to calculate a relative sensitivity factor (RSF) for that species in that matrix. A reference signal from the matrix (here, Nb) is used to adjust for instrumental factors. Example analyses of nitrogen implants in both poly and single crystalline material appear in Figure 2. RSF values, which are inversely proportional to the secondary ion yield, can then be used to convert secondary ion intensity to concentration using:

$$\rho_i = \left(\frac{I_i}{I_m}\right)(RSF)$$

Where $\rho_i$ is the impurity concentration in atoms/cm$^3$, $I_i$ is the measured impurity ion intensity, and $I_m$ is the measured matrix ion intensity. Most concentrations were reported in atomic ppm and denoted by ppm(a).

In order to accurately quantify data using an RSF value, the intensity of the matrix signal, in this case Nb, must stay relatively consistent from sample to sample. This is because the RSF value is calculated by normalizing the secondary ion signal to the matrix (Nb) signal. Early on in the development of a SIMS method it became clear the matrix signal was varying from implant standard to sample, and from sample to sample in some cases. Examination of mass spectra from a number of samples showed a number of samples had interferences caused by ions with fractional masses believed to be metastable hydrides. Figure 5 shows the mass spectra from two samples. The blue line shows a sample free of metastables with a Nb peak clearly resolved to baseline followed by several Nb hydride peaks also resolved to baseline. The orange line in Figure 5 shows the mass spectrum of a sample in which there is a large amount of metastable interference and the Nb hydride peaks are seen to be much larger. This phenomena was also observed by Maheshwari [23] and Stevie [24] and is discussed in more detail there.

It was shown by heat processing samples the metastable contribution to the secondary ion signal could be eliminated. The vacuum heat treatment now common for Nb cavities (800°C) lowers hydrogen levels in the Nb enough to avoid the formation of metastable hydrides.

## EXPERIMENTAL

### *Secondary Ion Mass Spectrometry*

One of the main differences among variations of SIMS instruments is the mass analyzer type. There are three that are common: quadrupole, magnetic-sector, and time of flight. Double focusing magnetic sector instruments are typically large and the most expensive. However, they have the ability to operate with high primary currents and have the highest sensitivity. This makes the dual focused magnetic sector instrument, such as the CAMECA IMS-7f GEO used here, the preferred choice for depth profiles and quantitative analyses of trace elements. Figure 6 shows a schematic of the Cameca 7f instrument.

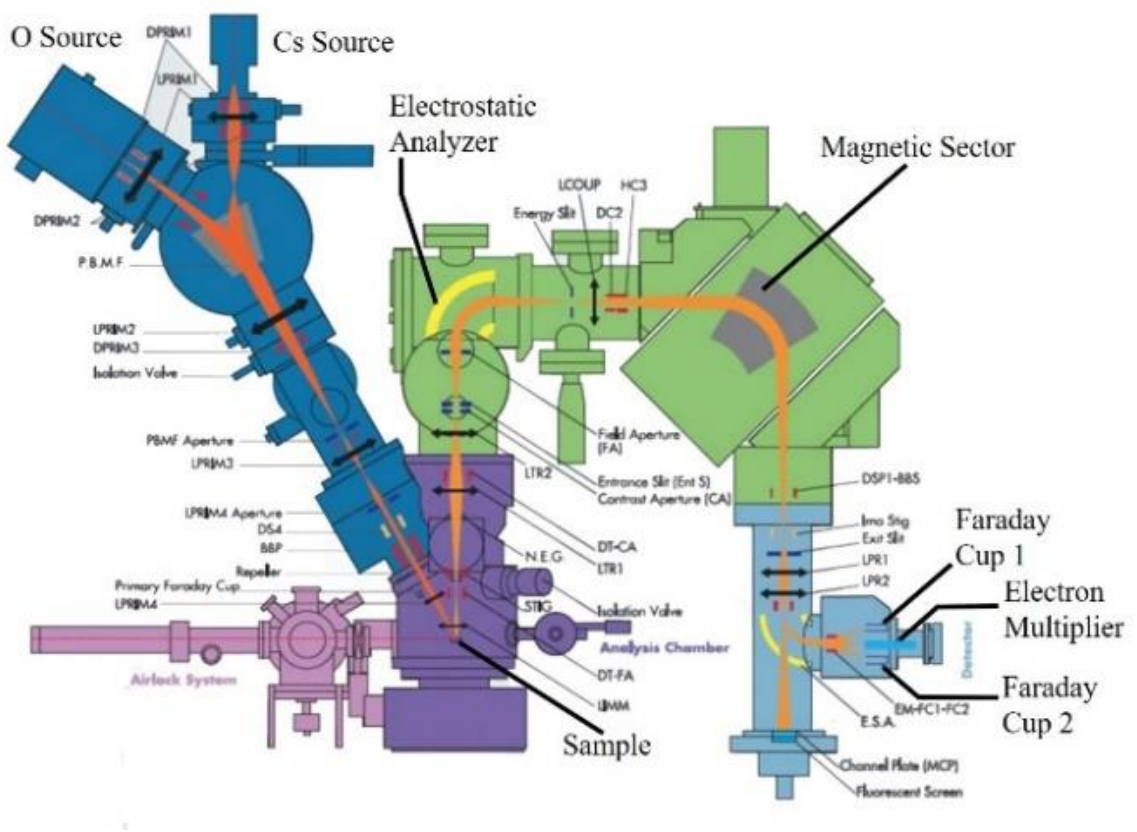

Figure 6: Schematic of Cameca IMS7f-GEO.

For nitrogen analysis, a $Cs^+$ primary ion beam was used with negative secondaries being detected. The species $^{93}Nb$ was used as a reference signal in all cases. Due to the fact that nitrogen negative ion yield is virtually zero under many analysis conditions, the molecular species ($^{93}Nb^{14}N^-$) was used for detection of nitrogen. Impact energy of 15 kV (10kV source/-5kV sample) was used with a current of 100nA rastered over a 150 × 150 µm area. A 63 µm diameter analysis area was used for typical depth profile measurements in order to cut crater edge effects.

When small spot analysis was needed, the Cs primary beam was tuned down to 5nA with a size of ~5 µm. The smaller beam was rastered over a 10 x 10 µm area. Due to the size of the beam in relation to the raster size, this gives an analysis spot size of ~15 µm, allowing analysis of single grains.

For analyses requiring higher depth resolution, such as low temperature addition samples, impact energy was lowered to 8 kV (+5 kV/-3 kV). While this maintained approximately the same angle of incidence as the 15 kV beam, 23.7° vs 24.4° respectively, the lower accelerating voltage lowered damage depth sufficiently to show a marked improvement in depth resolution. TRIM calculations show the majority of damage for the 15 kV condition to be ≤ 5.7 nm, while 8 kV is calculated as ≤ 4.0 nm. Empirically, using the interface method described above, the depth resolution increased ~10% from 6.2 nm to 5.6 nm. In addition to the lower impact energy positively affecting depth resolution due to physical effects, the lower sputter rate increases the data density making for a more accurate representation of the near surface region.

### *Sample Holder*

As mentioned, when analyzing for atmospherics at low concentration, good vacuum condition is paramount for low detection limits. A special sample holder for the Cameca 7f was designed and machined in order to analyze as many samples as possible at one time, thus reducing instrument exposure to atmosphere and normalizing the instrument conditions for up to twelve 6 ×6 mm sample coupons at one time.

### *Samples/Treatments*

Unless otherwise noted, witness samples were 10 mm square coupons cut by electrical discharge machining from trimmings of the 3 mm thick niobium sheet used to make SRF cavities ("RRR grade"). Typical grain size for polycrystalline material is in the 50 µm to 100 µm range in the un-annealed state, except for a few instances (as noted) of single or bi-crystals cut from large-grained ingot slices.

Coupons were prepared with differing surface conditions including: buffered chemical polishing (BCP), electropolishing (EP), and nanopolishing (NP).

Standards used for quantification were prepared by ion implantation with $^{14}N$ to a dose of $1\times10^{15}$ atoms/cm$^2$ at 160 keV and $^{16}O$ to a dose of $2\times10^{15}$ atoms/cm$^2$ at 180 keV. Treatments of standards, such as surface polish were matched as close as possible to that of the analyzed sample.

## RESULTS AND DISCUSSION

Provided here are several examples and brief discussion of SIMS being utilized to gain insight into SRF materials.

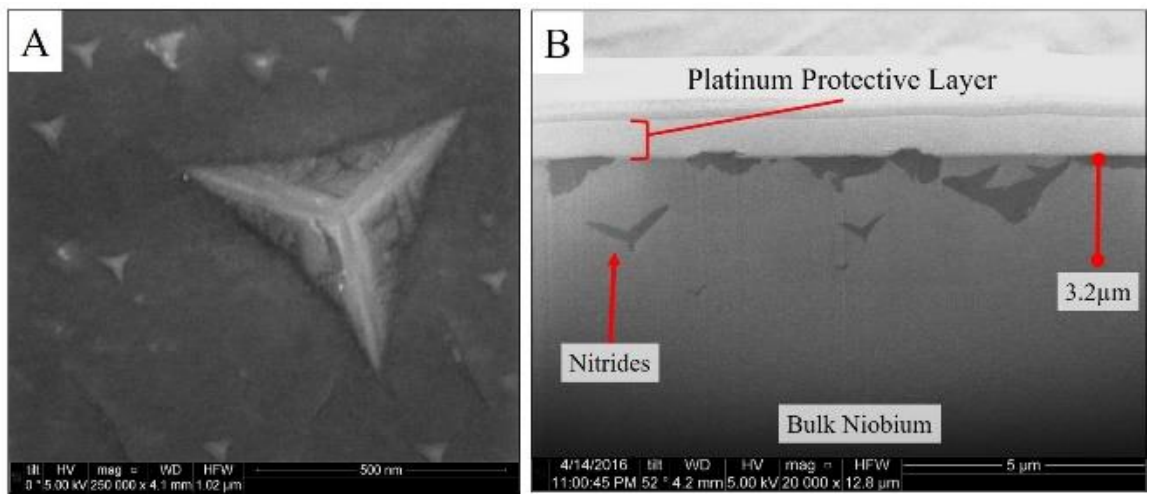

Figure 7: (A) Example of nitrides forming on surface of N-doped sample. (B) Shows a cross-section of a sample N-doped at 900 °C for 10 minutes at ~25 mTorr.

A lamentable side effect of the current N-doping recipe (800°C) is the formation of non-superconducting nitrides on the surface that can extend 3-4 microns in depth. (Fig. 7 A and B) This nitride layer must be removed, involving electropolishing (5 μm), ultrasonic degreasing, and high pressure rinsing. It would be beneficial, in regards to logistics and cost, to develop an alternate doping recipe which avoids nitride formation. While there are historical works dealing with nitridation kinetics in niobium, some even dealing directly with SRF cavities, the works take place at higher temperatures and pressures over relatively short "doping" times, making them not directly applicable. [1, 25, 3]

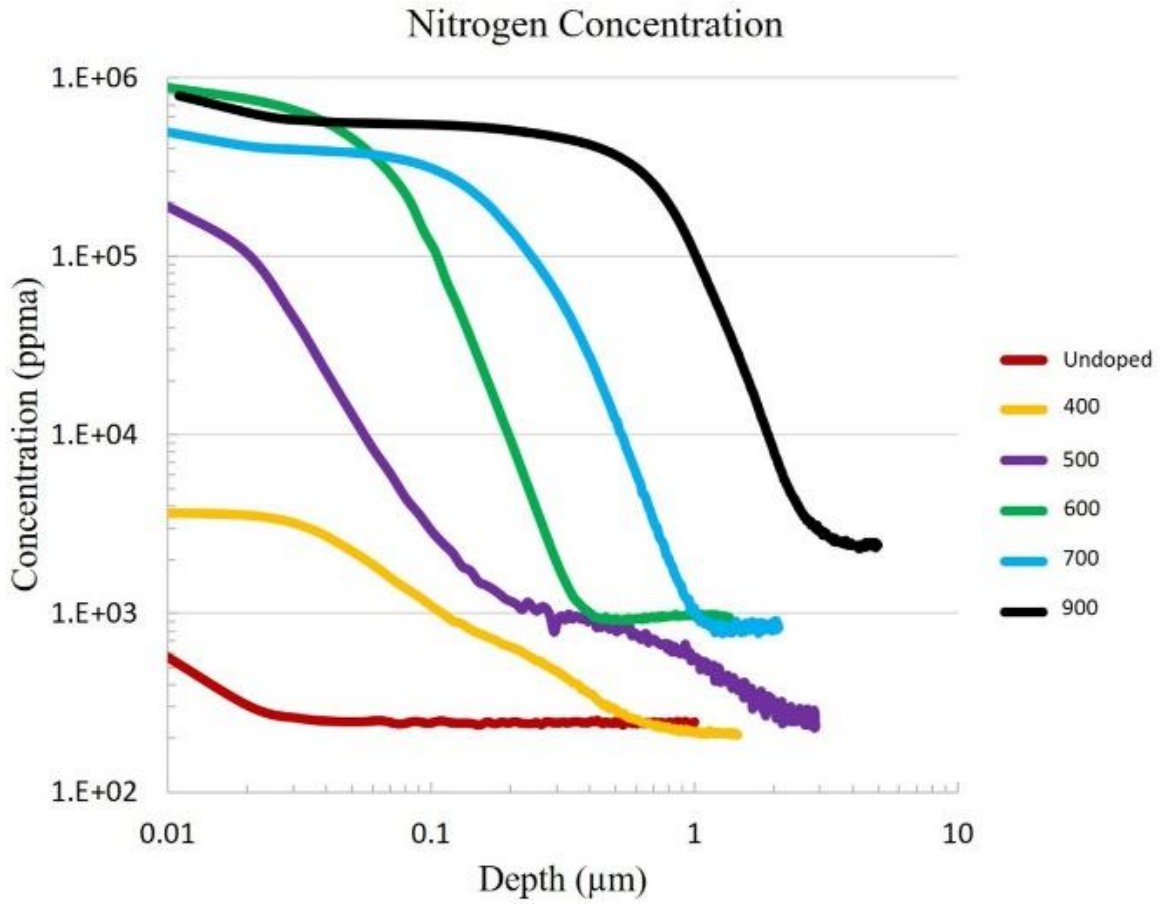

Figure 8: SIMS analyses of niobium samples doped at varying temperatures.

### *High Temperature Doping*

In order to investigate the possibility of nitride free doping on similar time scales as currently used, a set of NP samples were prepared varying the doping temperature from 900°C to 400°C. All samples were doped for the typical 20 mins, except for 900°C, which was doped for 10 mins. The samples were then analyzed by SIMS; conditions described in the experimental section. Figure 8 shows the resulting depth profiles. The red profile shows an undoped sample.There is a large step seen in the doped samples which represents the thickness of the nitride formed, while the last bit of each profile line is indicative of the amount of nitrogen doped into the bulk Nb. As the doping temperature is lowered, less nitride forms. Unfortunately, by the time the doping temperature is lowered to 400°C, nitride is formed and the nitrogen concentration is not raised. Indicating, at least on the time scale currently used, nitrogen impurity level cannot be raised without the formation of niobium nitride.

### *Low Temperature Doping*

Longer time scale, low temperature experiments are currently under investigation. [7, 26, 27, 6] Casually referred to as "nitrogen infusion", cavities show an increase in performance, similar, though not as drastic, as N-doped cavities. Here the process will be referred to as low temperature addition (LTA), as nitrogen appears not to be the sole player in performance gains.

As previously stated, SIMS instruments cannot be optimized simultaneously for depth resolution and sensitivity. [16] For bulk nitrogen measurements or N-doped coupon analysis, the highest priority for a SIMS analysis method is detection limit and proper quantification. For LTA, higher depth resolution is needed to properly describe the sample. SIMS was performed at the lower (8 kV) beam energy, with a depth resolution of 5.6 nm, described in the experimental section.

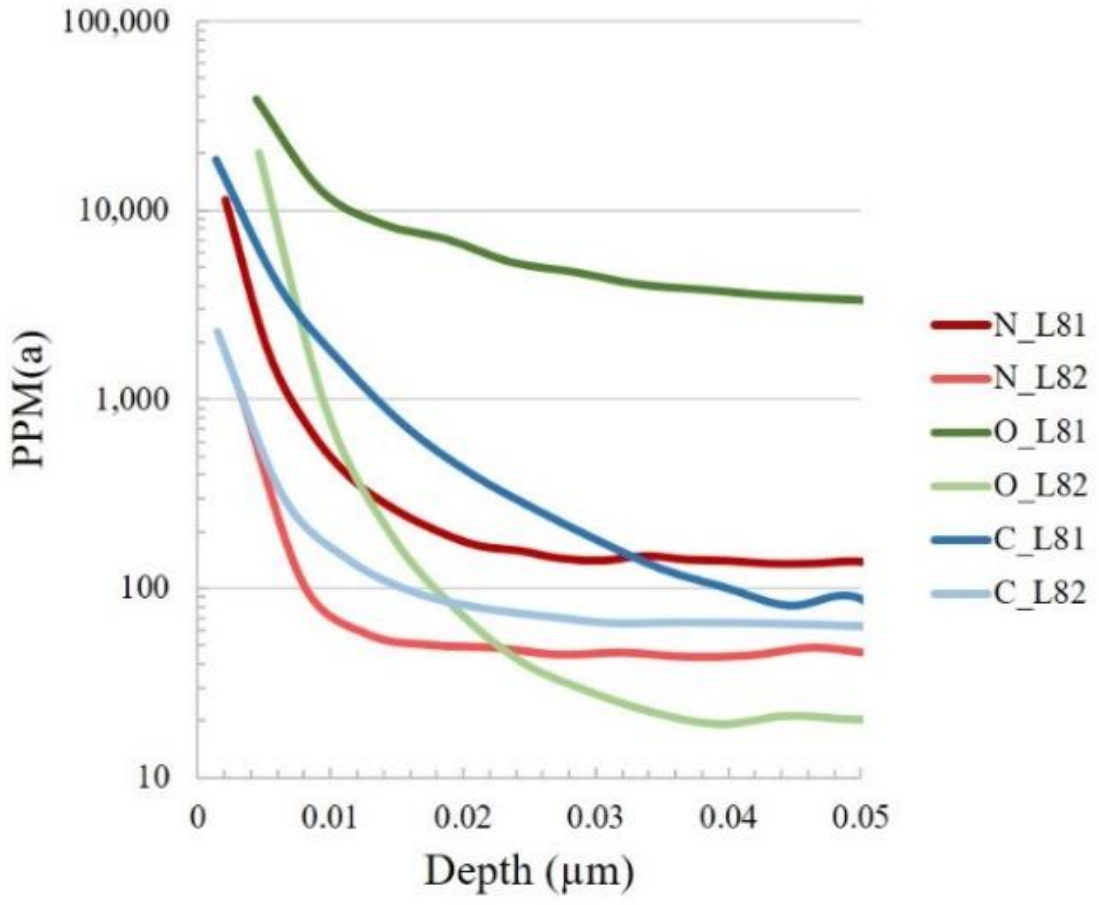

Figure 9: SIMS analysis of LTA sample (L81) and undoped (L82) sample in the near surface (first 50 nm).

Preliminary investigation was done using doped (L81) and undoped (L82) NP coupons. Sample L81 was held at 800°C for 3 hours, cooled to 140 $^{o}$C and then exposed to 25 mTorr $N_2$ for 48 hours during the same furnace run without exposing the sample to air. Several depth profiles for each sample were averaged together, results for the first 50 nm of the doped and undoped samples can be seen in Figure 9. The LTA technique is shown to raise all three impurity levels in the near surface region, having the most dramatic effect on the oxygen concentration.

### *As-Received Cavity Materials*

In this experiment, bulk niobium coupons received from three different suppliers and meant for cavity manufacture were analyzed by SIMS for nitrogen and oxygen content, establishing a baseline level for cavity raw material niobium for the first time. At the time of testing, different supplier material and production lots were showing variation in performance tests after N-doping. One possibility, differing amounts of impurities, most likely N and O, present pre-process. Material was used from three suppliers; Wah Chang, Tokyo Denkai, and Ningxia. Samples were marked W, T and N respectively, with numbers representing different lots of material.

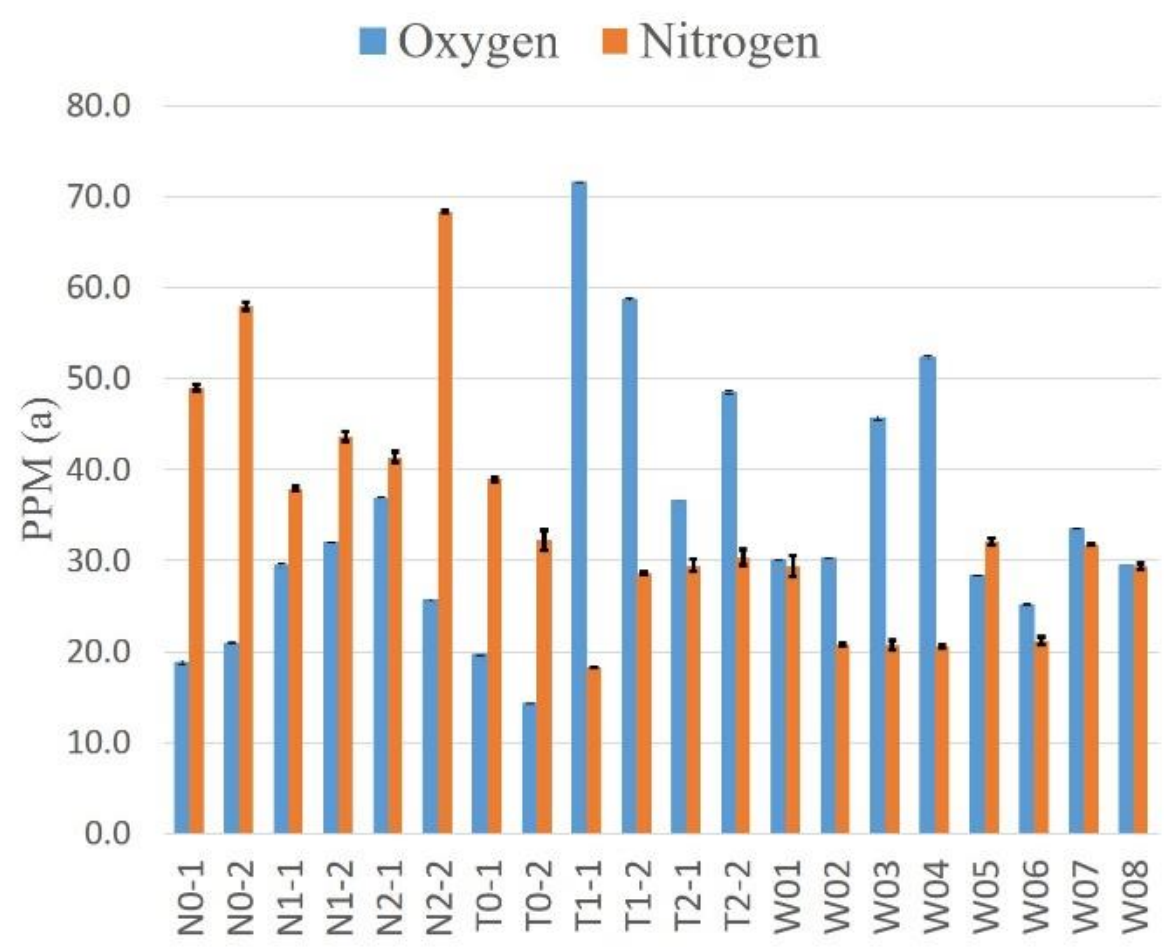

Figure 10: Oxygen and Nitrogen quantification with error for "as-received" samples.

Because of the number of samples, baseline analysis was split over several instrument loads spanning multiple days. With each load a raster reduction test was used to verify the nitrogen and oxygen signal was being detected within the sample and not background generated. Raster reductions for each load of samples appeared similar; Figure 4 shows an example raster reduction from this work. The raster reduction shows detected nitrogen originating from the sample, while part of the detected oxygen comes from instrument background. This indicates that while the quantification of nitrogen in samples is correct, actual oxygen levels are at or below the values reported here.

The analysis results can be found in Figure 10 along with the calculated error for each. As previously mentioned, crystal orientation can affect quantification. Here, error calculated between analysis regions on each sample are relatively small and representative bars are barely visible in Figure 10. This indicates that while grain size is on the same order of magnitude as the analyzed area (~63 µm), the crystal orientation from one analysis area to another does not cause significant variation in measurement. There is variation in nitrogen/oxygen concentration between manufacturers and lots. While error from the average crystal orientation differences between samples cannot be ruled out, total measured concentrations are small compared with doped samples. Results for nitrogen and oxygen in samples prepared to LCLS-II specs were found to be 1-2 ppth(a) and 300-500 ppm(a), respectively. Oxygen concentration across all as-received samples was found to be 34.4 ± 14.5 ppm(a), with a maximum of 71.6 ppm(a). As-received nitrogen concentration was found to be 34.1 ± 13.0 ppm(a) with a maximum value of 68.4 ppm(a). It may be noted that all but are less than the 10 wt ppm specification. Further, in all cases, the amount of nitrogen in as-received samples was found to be more than an order of magnitude lower than doped samples, indicating as-received values are not high enough to affect performance after doping.

### *N-Doping Orientation Dependency*

Previously, while analyzing bi-crystalline samples, a difference was observed from one side of the grain boundary to the other, giving rise to the question of whether grain orientation has an effect on the doping process. [28] SIMS was performed on each side of a central grain boundary within a few hundred microns of the boundary itself. A graphical representation of analysis placement can be seen in Fig. 11 along with optical images of the analysis craters.

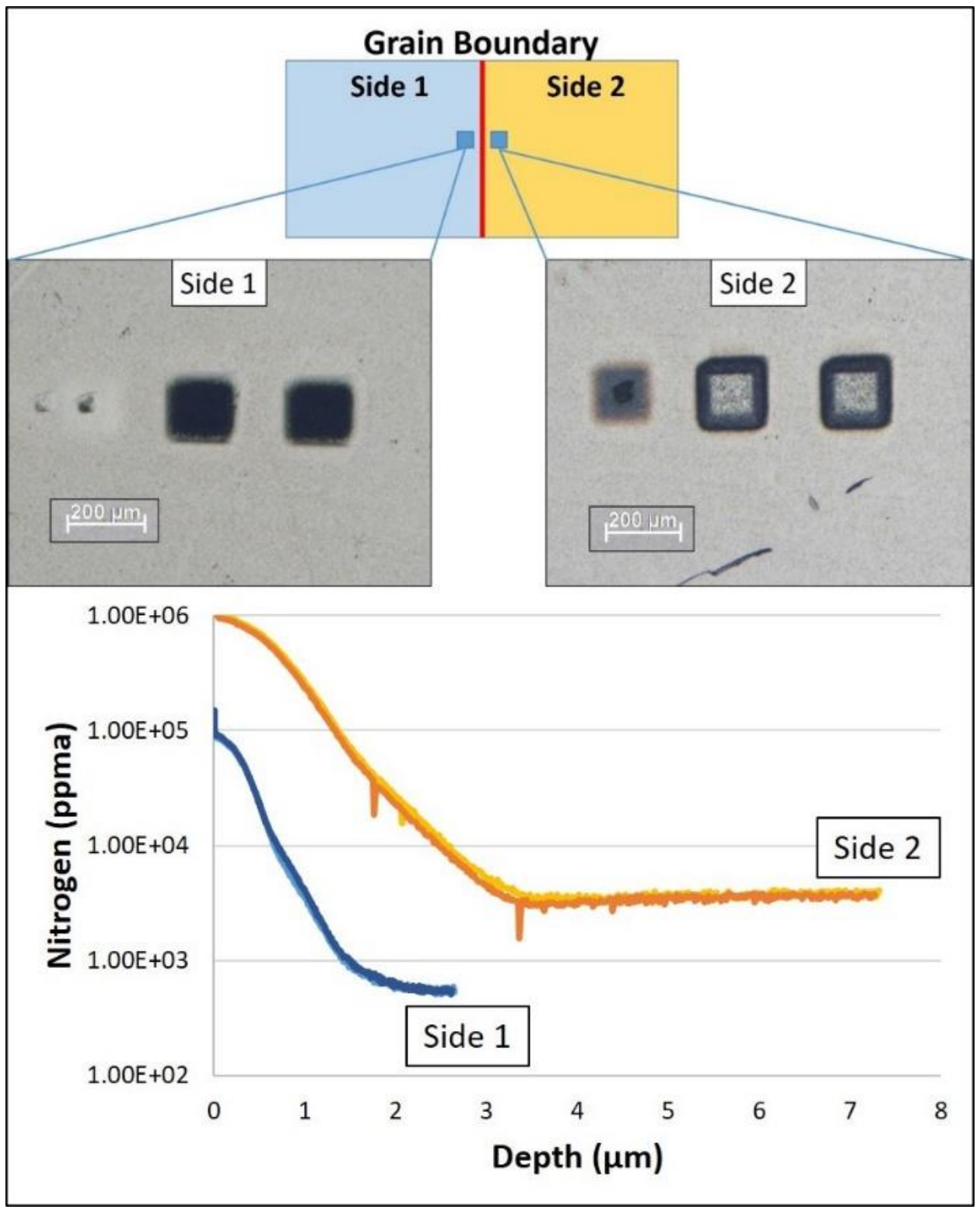

Figure 11: Illustration of crater placement, optical images of craters, and depth profile data from SIMS analysis of N-doped (900°C, 10 mins, ~25 Torr N) bicrystal sample.

Figure 11 also shows the depth profile data from all four craters. A clear difference can be seen in the profile from one side of the grain boundary to the other. As crystal orientation is the only differentiating characteristic between the analysis points, it must be tied to the differences seen in the data, either through instrumental effects, such as differences in $S_{sec}$ or differences in the sample, such as doped nitrogen concentration.

Crystal orientation is known to affect sputter rate during SIMS analysis. The relative ion yield may also be affected causing differences in quantification. [20] In addition, niobium is a body centered cubic structure, which should not exhibit diffusion rate differences based on orientation. This suggests that the differences seen in nitrogen concentration in the bicrystalline sample may be instrumental related rather than due to the orientation having some effect on N-doping. However, the niobium nitride formed on the surface during doping plays a vital role in the uptake of nitrogen into the bulk niobium [25], and the orientation dependence for the formation of this nitride can be clearly seen. (Fig. 12 A and B)

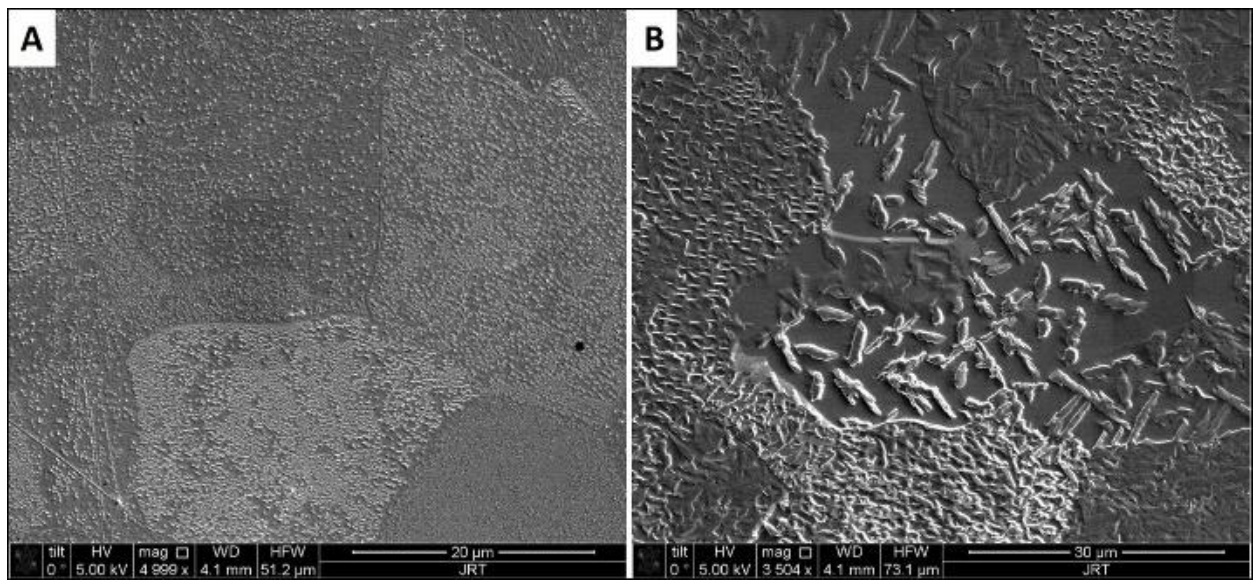

Figure 12: Image showing surface of 700°C N-doped sample (A) and lightly sputtered (focused ion beam) area of 900°C N-doped sample. (B) Both show clear differences in nitride formation from grain to grain.

In order to further investigate, bi-crystalline samples prepared as part of the high temperature doping study were analyzed. Figure 13 shows SIMS results from the doped bi-crystalline samples. X-axis shows doping temperature while Y-axis shows measured nitrogen concentration in the bulk. After SIMS analysis, electron backscatter diffraction (EBSD) was used to determine the orientation of each analysis crater. Larger differences were seen in orientation in samples that showed larger differences in N concentration. Figure 14 shows inverse pole figures for both grains overlaid for the 700°C and 900°C samples. In contrast to the 700°C sample, the 900°C doped sample shows two distinct orientations between grain A and B. Of note is the 400°C sample, which shows a larger variation in orientation than the 500°C or 700°C samples, but shows less difference in measured N concentration.

RSF values calculated from bi-crystalline and single crystal implant standards have been seen to vary from 10-50% between orientations. However, since implant standards were not available for each possible crystal orientation, an average RSF value was used to quantify the data here. The average RSF, was calculated based on data from multiple bi-crystal implant standards and collected using the same instrument conditions (10kV/-5kV/100nA).

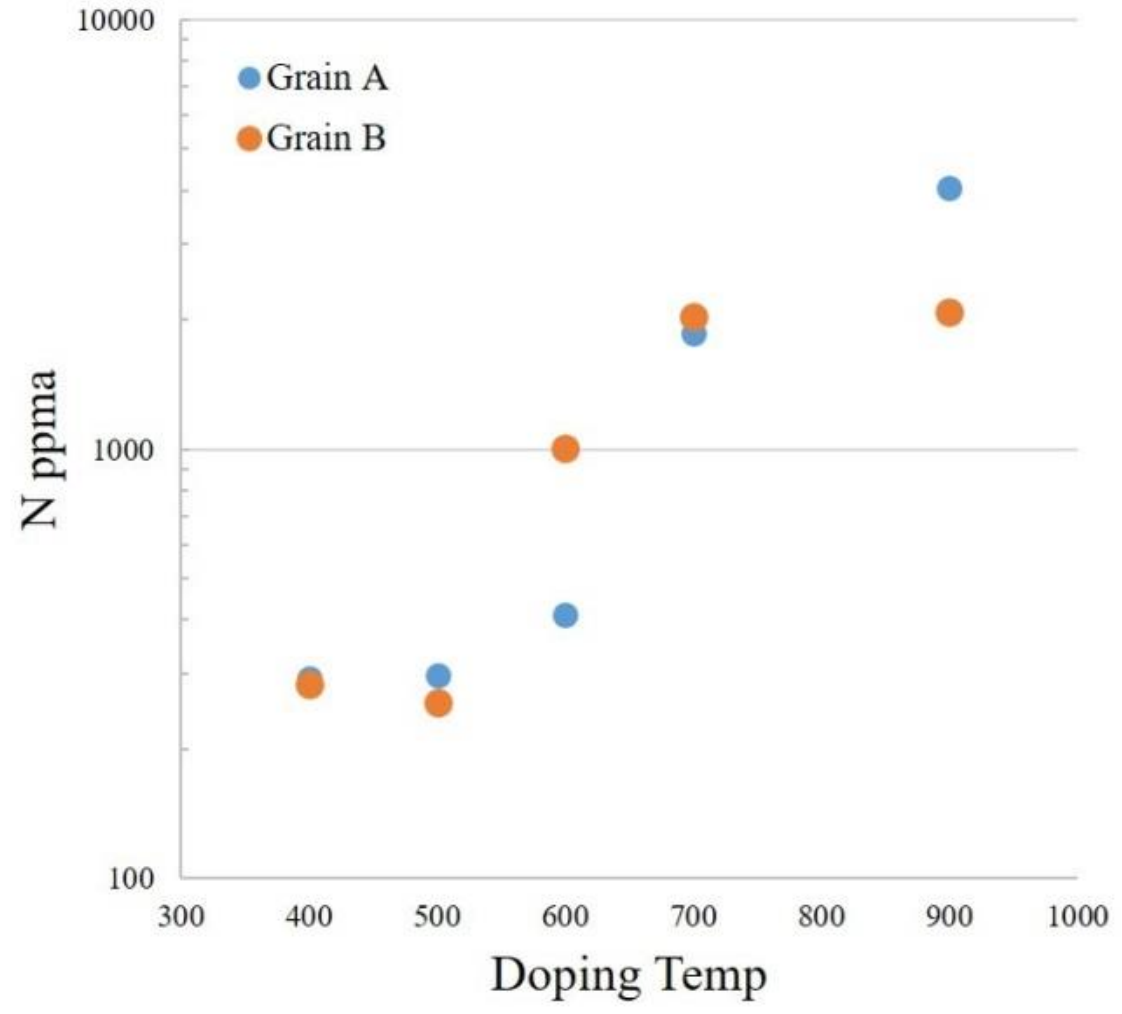

Figure 13: SIMS results from bi-crystalline samples doped at different temperatures. Grain A and B represent two sides of a grain boundary.

While there are significant sources of error to be corrected, evidence seems to point to an orientation dependence for N-doping. Future work could include the meticulous task of creating N-doped and implant standards from the same large grains in order to eliminate uncertainty.

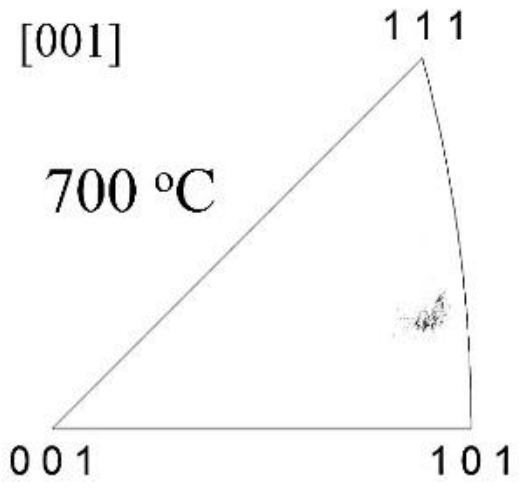

Figure 14: Inverse pole figures showing overlays of data collected from grains A and B for the 700°C and 900°C bi-crystalline samples.

## *$Nb_3Sn$ Contamination*

The discussion above is about transitioning a demonstrated successful technology (doping) from the research stage to reliably successful deployment: from the state of the art to the state of the practice. Coating with $Nb_3Sn$ has demonstrated attractive potential for more than forty years but has still to demonstrate complete success. The chief shortcoming continues to be unacceptable decline of cavity quality factor with increasing gradient. There has yet to emerge any consensus about the cause.

The ability of SIMS to detect very low concentrations of impurities led us to search for some possible contaminant present in nearly all experiments. Here we consider titanium. A possible source could be the Nb/Ti flanges widely used for their superior mechanical strength. Cornell coated cavities, which tend to have little $Q_0$ slope compared to those coated elsewhere, use Nb flanges. In this study, several witness coupons were coated at Jefferson Lab. Figure 15 shows SIMS depth profiles from single crystal and polycrystalline witness coupons coated with $Nb_3Sn$ under cavity coating conditions. In this case, only the absence or presence of Ti is of concern, and the concentration is not quantified, but represented in counts. The depth scale is quantified as the location of the Ti is of interest.

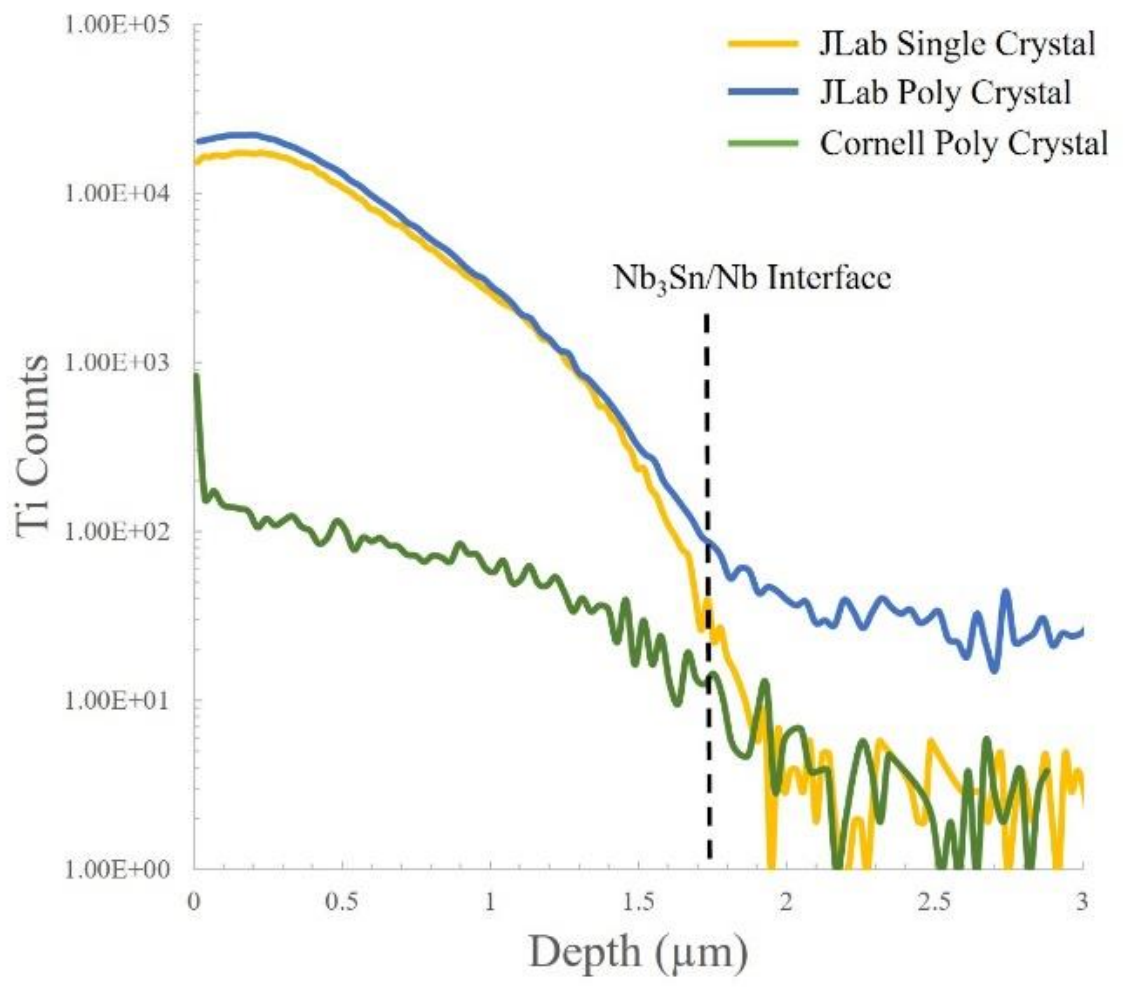

Figure 15: SIMS depth profile of Ti in $Nb_3Sn$ coated coupons.

In the case of the single crystal coupon, Ti is seen throughout the thickness of the $Nb_3Sn$ layer, but then quickly falls to the detection limit. When the same experiment is conducted using a polycrystalline witness sample, the Ti signal stays a full decade above the detection limit past the $Nb_3Sn$ layer. Raster reduction method and mass spectra were used to confirm the presence or absence of Ti in the bulk Nb under the $Nb_3Sn$ coating. The difference is possibly due to migration of Ti into the bulk Nb via grain boundaries not available in the single crystal depth profile.

After verifying some test cavities coated at Jefferson Lab contained a “significant” amount of Ti, a Jefferson Lab manufactured cavity was transported and coated at the Cornell facility. Witness coupons were coated along with the cavity. Figure 15 also shows the SIMS depth profile of the Cornell witness sample. While the Ti signal has not been quantified, it has been normalized based on the matrix signal, and by comparing the Ti signal difference between the Cornell and Jefferson Lab witness samples, it is clear the Jefferson Lab coated samples contain a considerably greater amount of Ti. Currently cavity performance testing is underway to resolve whether these differences are causative of cavity performance affects.

## CONCLUSION

There are several materials-based technologies being developed that promise to push SRF performance forward, resulting in higher performing and more efficient particle accelerators. In order to push such technologies forward in a timely and efficient manner, accurate and robust materials characterization techniques are needed, linking a material's structure to processing and performance. SIMS has proven vital in the goal of linking processing, structure, and performance.

## ACKNOWLEDGEMENT

- Co-Authored by Jefferson Science Associates, LLC under U.S. DOE Contract No. DE-AC05-06OR23177
- We are grateful for support from the Office of High Energy Physics, U.S. Department of Energy under grant DE-SC0014475 to the College of William & Mary.
- The Institute for Critical Technology and Applied Science (ICTAS), and Nanoscale Characterization and Fabrication Laboratory (NCFL) at Virginia Tech
- Stanford Linac Coherent Light Source II (LCLS_II) for supplying the Ningxia and Tokyo Denkai production material samples.
- The SRF research group at Cornell University for coating certain of our materials